# Laser-induced damage thresholds of ultrathin targets and their constrain on laser contrast in laser-driven ion acceleration experiments

Dahui Wang[a,b], Yinren Shou[a], Pengjie Wang[a], Jianbo Liu[a], Zhusong Mei[a], Zhengxuan Cao[a], Jianmin Zhang[b], Pengling Yang[b], Guobin Feng[b], Shiyou Chen[a], Yanying Zhao[a], Joerg Schreiber[c], Wenjun Ma[a,*]

[a]State Key Laboratory of Nuclear Physics and Technology, Peking University, Beijing, 100871, China

[b]State Key Laborartory of Laser Interaction with Matter, Northwest Institute of Nuclear Technology, Xi'an, 710024, China

[c]FakultätfürPhysik, Ludwig-Maximilians-University, Am Coulombwall 1, D-85748 Garching, Germany

Single-shot laser-induced damage threshold (LIDT) measurements of multi-type free-standing ultrathin foils were performed in vacuum environment for 800 nm laser pulses with durations τ ranging from 50 fs to 200 ps. Results show that the laser damage threshold fluences (DTFs) of the ultrathin foils are significantly lower than those of corresponding bulk materials. Wide band gap dielectric targets such as SiN and formvar have larger DTFs than those of semiconductive and conductive targets by 1-3 orders of magnitude depending on the pulse duration. The damage mechanisms for different types of targets are studied. Based on the measurement, the constrain of the LIDTs on the laser contrast is discussed.

## 1. Introduction

The application of chirped-pulse amplification (CPA) in solid lasers has realized the output of lasers with femtosecond duration and petawatt-class power [1-2]. Interaction of such ultra-intense laser pulses with thin foils has generated protons up to 100 MeV and carbon ions close to 600 MeV [3-8], which has potential applications in the fields of hadron therapy, fast-ignition laser fusion, isotope production and proton radiography [9-11]. Especially, when the thicknesses of the foils are in the range of several nanometers to a few tens of nanometers, quasi-monoenergetic ions can be generated in the scheme of radiation pressure acceleration (RPA) [12-14]. When ultrathin targets are used, the damage caused by the prepulse energy should be avoided. For a high-power laser pulse with a contrast of $10^7 \sim 10^9$ and intensity of $\sim 10^{20}$ W/cm$^2$, the prepulse intensity of ASE or prepulses is in the range of $10^{11}$-$10^{13}$ W/cm$^2$, which is high enough to damage many targets and significantly influence the acceleration process [15-17]. For the upcoming 10 PW class laser facilities such as ELI [18-19], the prepulse intensity will scale up with the increasing peak power. Hence, the choice of target material which can survive the prepulse energy would be a crucial issue for laser-driven ion acceleration.

Ultrathin free-standing foils made of different materials, such as metal, graphite, diamond-like carbon (DLC), and transparent polymers, have been employed as targets in laser ion acceleration experiments for years. I. J. Kim *et al.* reported the generation of 93 MeV protons by using 15 nm formvar, a kind of transparent polymer, as targets irradiated by a PW laser [3]. I. Prencipe *et al.* reviewed the ultrathin targets used for ion acceleration and the corresponding fabrication techniques, emphasizing that ultrahigh laser contrast is pre-requisite for this application [20]. The contrast of the laser pulses in many cases imposes a substantial restriction on the choice of targets. Therefore, the laser-induced damage thresholds (LIDTs) of the ultrathin targets are essential parameters that should be learned before the acceleration experiments. Generally, LIDTs of the ultrathin foils are considered the same as that of corresponding bulk materials. As a matter of fact, they could be significantly different. For nanometer-thin metal foils, their thicknesses are much smaller than their heat deposition depths. The mean energy density in the foils could be significantly higher than that in corresponding bulk materials. For ultrathin dielectric foils, the existence of massive surface defects induces surface states within the bandgap and intrinsic seed carriers, which will affect the ionization and laser energy transport. It is found that, due to the existence of the surface deffects, the LIDTs of 5nm-40nm DLC foils are significantly smaller than that of bulk DLC[21]. Besides of DLC foils, systematic studies on the single-shot LIDTs of other type of free-standing ultrathin targets were still missing.

In this letter, we present a study on the single-shot LIDTs of multi-type free-standing ultrathin foils for a laser pulse with wavelength of 800 nm and pulse durations from 50 fs to 200 ps. Widely used targets made of formvar, silicon nitride (SiN), aluminum (Al), amorphous carbon (a-C), and carbon nanotubes (CNT) were tested. Our results show that the damage thresholds of laser fluence (DTFs) of all the ultrathin foils are significantly lower than that of corresponding bulk materials. Formvar and SiN foils have the highest DTF among the tested targets. Their DTFs scale up with the laser pulse duration. For Al, amorphous carbon and CNT foils, the DTFs are weekly dependent on pulse duration in general. By comparing the damage intensity of the targets with the prepulse intensity of a laser system, the constrain of targets' LIDTs on the laser contrast can be obtained in a simple way.

2. **Experimental setup**

The experiments were performed on 100s TW Ti:Sapphire laser systems of CLAPA in Peking university and on ATLAS in Max-Planck-Institutfür Quantum Optik. Seed pulses from a Kerr-lens mode-lock oscillator were amplified in a linear regenerative amplifier followed by a booster amplifier. The laser pulses were focused onto the samples by an f/3.5 off-axis parabola (OAP) mirror, which was used to generate energetic ions or x/γ ray emission in other experiments[22]. The full width of half maximum (FWHM) diameter of the focal spot is 4.5μm, which is measured by a microscopic imaging system equipped with a 12 bit charged coupled device (CCD). The laser energy was controlled by using different neutral density optical filters before the compressor. All the filters were thinner than 2 mm, avoiding the effects of self-phase modulation and wavefront distortions. A pyroelectric detector PE50 (OPHIR) was used to measure the energy and calibrate the filters. The RMS energy stability of the laser system was typically less than 2%, and we report the average value here. The pulse duration, ranging from 50 fs to 5 ps, was adjusted by moving the diffraction grating in the compressor chamber and measured by a dispersion-minimized autocorrelator (Amplitude). Pulses with durations of 200 ps were obtained by by passing the compressor.

The foils were positioned precisely at the focal spot controlled by a high-precision position system. After each shot, the foils were front-illuminated by a collimated light source. A long working-distance microscope objective lens was driven in behind the foils to image the irradiated foils. The images were recorded by the CCD behind the lens with a high resolution[22]. The damaged area of the foils is defined by the permanent holes induced by the laser pulses on the targets. The smallest damage spot we could resolve was approximately 0.5 μm. All the measurements were performed in vacuum environment with the pressure of 3×10$^{-3}$ Pa.

## 3. Methods and results

The LIDTs of 50-nm formvar, SiN, Al, a-C foils, and 50-μm free-standing CNT foams. were measured in the experiments. They were fabricated by the methods of spin-coating(formvar), plasma-enhanced chemical vapor deposition(SiN), thermal deposition (Al, a-C) and chemical vapor deposition (CNT), respectively[20, 23]. The carbon nanotube targets were used as near-critical-density targets in previous studies[4, 7]. Their average density here is 4.5 mg/cm$^3$. The areal density of a 50 μm CNT target is equal to an 80 nm DLC target.

In the experiments, single-shot measurements of the LIDT were performed in order to avoid any incubation or fatigue effects induced by multiple pulses. Fig.1(a)(b) show the damage morphology of a 40 nm formvar foil irradiated by a pulse with full-width-at-half-maximum (FWHM) duration of 50 fs at the intensity of 6×10$^{12}$ W/cm$^2$ and 1.1×10$^{13}$ W/cm$^2$, respectively. The intensity distribution around the focal spot of the laser is shown in Fig.1(c). The damaged areas of the targets are centered at the peak of the laser pulses, with shapes following the spatial distribution of the laser intensity. It indicates that the damage is intensity-determined. Collateral damage due to thermal diffusion is not dominant.

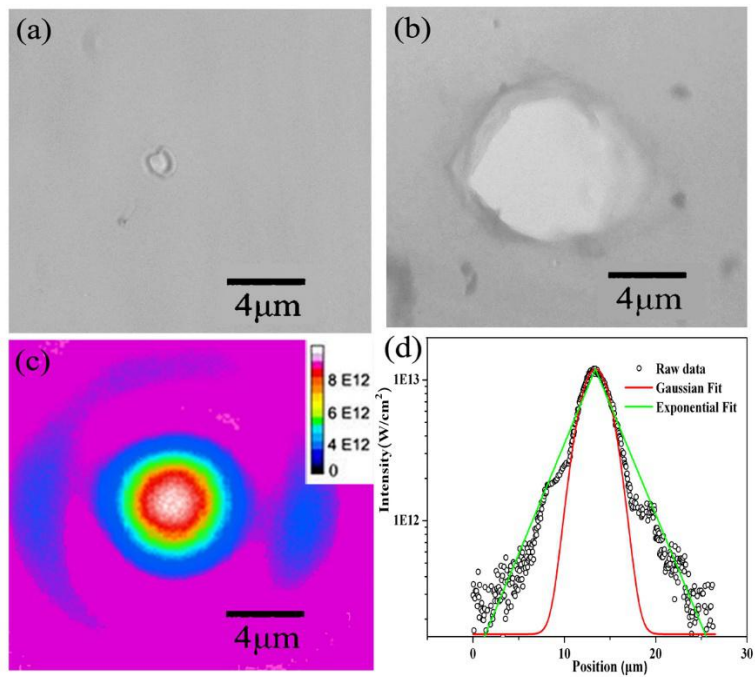

Fig.1 The damage morphologies of a formvar target under laser irradiation at the intensity of 6×10$^{12}$ W/cm$^2$ (a) and 1.1×10$^{13}$ W/cm$^2$(b). (c) The intensity distribution around the focal spot. (d) A line out of the laser intensity across the focal spot (scattered black circle), the corresponding Gaussian fit (red line), and the exponential fit (green line) of the raw data.

The LIDTs can be obtained by analyzing the damage morphologies at different laser intensity. Fig.2 shows the damage diameter of D with respect to the peak intensity of *I*. One can see the data can be fit by $D \propto lnI$[21], different from the results of $D^2 \propto lnI$ when low-power lasers are employed[24-25]. This is because, when the focal intensity is much higher than the damage threshold, the diameter of the damaged area is determined by the intensity far away from the focal spot. For 100s TW lasers, the intensity distribution $3D_{FWHM}$ away from the focal spot typically has an exponential distribution as shown in Fig.1(d) due to the wavefront distortion instead of a Gaussian distribution. Such distortion will not be removed by adding neutral density filters. So the intensity distribution around the focal spot is the same for attenuated beam. By extrapolating the data to $D=0$ in Fig.2, we obtain $I_{th}$=4.5×10$^{12}$ W/cm$^2$, which is the single-shot damage threshold of laser intensity (DTI) for the 50 nm formvar target at $\tau$=50 fs. Then, the corresponding single-shot damage threshold of laser fluence (DTF) $F_{th}$ can be derived as $F_{th} = I_{th} \times \tau$ =0.18 J/cm$^2$.

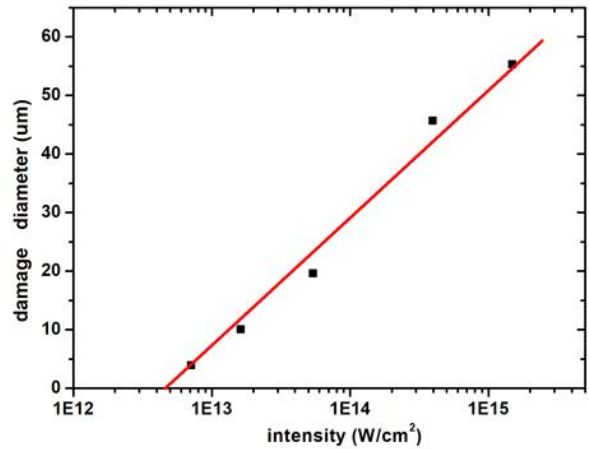

Fig.2 The diameters of the damaged area with respect to the laser intensity

Using the same method, we measured the LIDTs of other ultrathin foils. Fig.3 shows the DTFs with respect to the pulse durations ranging from 50 fs to 200 ps. For a systematic comparison, the result of 40 nm DLC foils obtained in previous study is also included. Generally, compared to the DTFs of micrometer-thick foils, the DTFs for these ultrathin foils are smaller by order of 1-2 [25-27]. DTFs of formvar and SiN foils are significantly higher than those of other foils, showing apparent dependence on $\tau$.

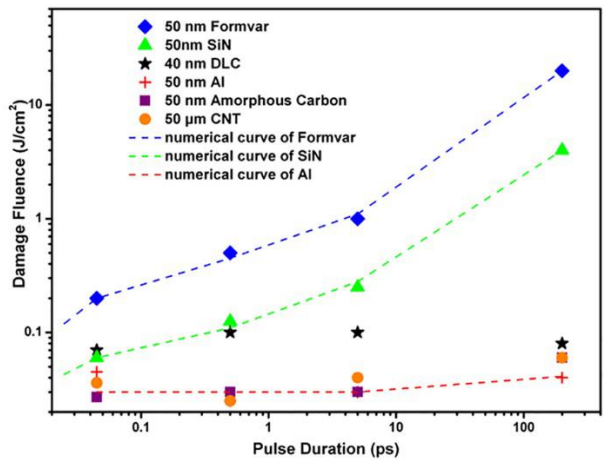

Fig.3 Pulse duration dependence of DTF for Formvar (bule diamond), SiN (green triangle), DLC (black star), Al (red cross), amorphous carbon (purple square) and CNT (orange circle)

# 4. Analysis of the damage mechanisms of different targets

According to their conductivity, the targets in this study can be categorized into three categories: formvar and SiN are the insulators, DLC is a semiconductor, while Al, amorphous carbon, and CNTs are the conductors. For insulators and semiconductors, free electrons/holes are generated through multiphoton absorption and avalanche ionization. Irradiated by the femtosecond laser pulses, damage occurs when the density of free electrons reaches the critical density $n_{cr}$ ($n_{cr} = 1.1 \times 10^{21}/\lambda^2$ cm$^{-3}$). The conductors have free electrons absorbing the laser energy. Damage occurs when the absorbed energy per atom exceeds the bonding energy. Next, we will make the analysis of LIDTs results and discuss the damage mechanisms of different targets.

## 4.1 Formvar and SiN

As shown in Fig.3, DTFs of formvar are higher than that of SiN for all the pulse durations. The dependences of DTFs on the pulse durations for formvar and SiN foils are similar. The DTFs of both foils scale up with $\tau^{1/2}$ (actual fit in our cases: $\tau^{0.51}$) for $\tau > 10$ ps. When $\tau < 10$ ps, the dependence deviates from this scaling.

For long pulses ($\tau > 10$ ps), as discussed in previous literature[28-33], the laser intensity is not high enough to initiate an avalanche ionization originating from seed electrons. The damage is due to the fracture or melting caused by the high laser fluence. The rate-limiting process for the rise of temperature is lattice thermal diffusion, and the DTF is proportional to $\tau^{1/2}$. For pulses shorter than 10 ps, the energy absorbed from the laser field can not be transferred to the lattice as fast as it is deposited in the electrons. The rate equation of carriers [28-30] can be utilized to analyze the generation and evolution of the free electron density $n(t)$ as follows

$$\frac{dn(t)}{dt} = \delta I(t) n(t) + P(I) \tag{1}$$

Here, $\delta$ is the avalanche coefficient. $P(I) = \sigma_k (I/\hbar\omega)^k N_s$ with $k$ the multiphoton order, $\sigma_k$ the $k_{th}$ absorption cross-section and $N_s$ the atom density of the foil. If we completely separate the multiphoton and avalanche ionization processes, the DTF can be expressed as $\emptyset_{cr} \propto \ln(\frac{n_{cr}}{n_0})$, where $n_{cr}$ ($\approx 10^{21} cm^{-3}$) is the critical electron density[30], $n_0$ is the density of the seed electrons for avalanche ionization resulted from multiphoton ionization or defects in the foil. If the number of seed electrons was independent of intensity, e.g., as the result of defects, $\emptyset_{cr}$ would be independent of the pulse duration. In the cases of formvar and SiN foil, $\emptyset_{cr}$ rapidly decreases with pulse duration, it indicates multiphoton ionization plays an important role in the ultrashort pulse limit. Actually, the initial electrons ionized by photons can be calculated as[29]

$$n_0 = \int_{-\infty}^{+\infty} P(I)\, dt = N_s \int_{-\infty}^{+\infty} \sigma_k (I/\hbar\omega)^k \, dt = \sigma_k N_s (I/\hbar\omega)^k (\pi/\ln 2)^{1/2} \tau/k. \tag{2}$$

Set the smallest $k$ to satisfy the equation $k\hbar\omega \geq \Delta$, where $\hbar\omega = 1.55\ eV$, $\Delta_{formvar} = 5.5\ eV$ and $\Delta_{SiN} = 4.5\ eV$ are the photon energy of the 800 nm laser, the band gap energies of formvar and SiN. We can get the corresponding $k$ to be 4 and 3 respectively. Thus, the multiphoton ionization rate scales with intensity as $I^4$ and $I^3$ for formvar and SiN. For the laser fluence $\emptyset = \int_0^\tau I(t)dt$ and the Gaussian distribution in the time domain, the DTFs of SiN and formvar under the irradiation of ultrashort pulse are given by

$$\emptyset_{cr-SiN} = \frac{2}{\alpha}\ln\left(\frac{n_{cr}}{n_{0-SiN}}\right) = \frac{2}{\alpha}\ln n_{cr} - \frac{2}{\alpha}\ln\frac{\sigma_3 N_s}{3}\left(\frac{\emptyset_{cr-SiN}}{\hbar\omega}\right)^3 (\pi/\ln 2)^{1/2} - \frac{4}{\alpha}\ln\tau \tag{3}$$

and

$$\emptyset_{cr-formvar} = \frac{2}{\alpha}\ln\left(\frac{n_{cr}}{n_{0-formvar}}\right) = \frac{2}{\alpha}\ln n_{cr} - \frac{2}{\alpha}\ln\frac{\sigma_4 N_s}{4}\left(\frac{\emptyset_{cr-formvar}}{\hbar\omega}\right)^4 (\pi/\ln 2)^{1/2} - \frac{6}{\alpha}\ln\tau \qquad (4)$$

One can infer that the larger bandgap of formvar results in higher $\emptyset_{cr}$. Besides, a deviation from $\tau^{1/2}$ scaling can be observed in Eq.(3) and Eq.(4). The dashed lines in Fig.3 shows the simulated DTF for formvar and SiN by solving Eq (3) and Eq (4). They indicate that the classical model can explain the experimental results.

*4.2 DLC*

The bandgap of DLC is 3.5 eV, which distinguishes it from formvar and SiN as a semiconductor. The measurement on DLC foils shows that their DTFs are weakly dependent on the pulse duration, which is quite different formvar and SiN. For ultrathin DLC targets fabricated by cathodic arc deposition, a large number of defects and impurities results in high initial seed electron density (ISED) [34-36]. Because of the high ISED, avalanche ionization can be easily triggered even without enough contribution from multiphoton ionization. As a result, the DTF is determined by $F_{th} \propto \ln\left(\frac{n_{cr}}{n_0}\right)$, independent of the pulse duration.

*4.3 Al, amorphous carbon and CNT*

For these three kinds of conductive materials, the values and the dependency on the pulse duration of DTF are similar. For simplicity, we choose Al foils for investigation. Laser energy deposition and transport inside the metal can be described by the two-temperature model [37-38]

$$C_e \frac{\partial T_e}{\partial t} = \frac{\partial}{\partial x} \kappa \frac{\partial}{\partial x} T_e - g(T_e - T_i) + A(x,t), \qquad (5)$$

and

$$C_i \frac{\partial}{\partial t} T_i = g(T_e - T_i). \qquad (6)$$

Here, the electron heat capacity $C_e$ is given by $C_e = C_e' T_e$, $C_e'$ being a constant. κ is the heat conductivity, $T_e$ and $T_i$ are is the temperatures of the electrons and lattice, respectively. $C_i$ is the lattice heat capacity. $A(x,t)$ is the source term and $g$ is the electron-photon coupling constant.

Assuming that heat is deposited into the electrons on the surface and neglecting the initial electron temperature, we have the heat deposition depth[38]

$$x_R = \left[\frac{128}{\pi}\right]^{1/8}\left[\frac{\kappa_0^2 C_i}{T_{im}g^2 C_e}\right]^{1/4}, \qquad (7)$$

where $T_{im}$ is the melting temperature For Al, $C_i = 2 \times 10^6$ J.m$^{-3}$.K$^{-1}$, $C_e' = 91.2$ J.m$^{-3}$.K$^{-2}$, $T_{im} = 933$ K, $g = 2.45 \times 10^{17}$ W.m$^{-3}$.K$^{-1}$, $\kappa_0 = 238$ W/(m.K)[39]. The $x_R$ can be calculated to be 3.4 µm. This value is much larger than the thickness of the foils. Thus, if the pulse duration is shorter than the critical time $\tau_c$, the absorbed heat can be viewed as uniformly distributed on all the atoms in the foil, and the DFT is independent of the pulse duration. The critical time of $\tau_c$ can be estimated by assuming $T_e = T_i = T$, $C_e \ll C_i$ in Eq.(5) and Eq.(6)[24]. It can be obtained that

$$C_i \frac{\partial}{\partial t} T \approx \kappa \frac{\partial^2}{\partial x^2} T + A(x,t), \tag{8}$$

and

$$\tau_c = \frac{1}{2} C_i x_R^2 / \kappa_0. \tag{9}$$

From Eq.(7) and Eq.(9), the critical time can be derived as

$$\tau_c = (8/\pi)^{1/4} (C_i^3 / C_e' T_{im})^{1/2} / g. \tag{10}$$

For Al foils, the calculated $\tau_c$ is 5.6 ps. This value is in good agreement with the experimental results that DFTs are independent on pulse durations for $\tau < 10$ ps. If $\tau$ is longer than $\tau_c$, the diffusion of the lattice becomes dominated and DTF of the foils will scale as $E_{\text{th}} \propto \tau^{1/2}$, which is similar to the case of an insulator.

## 5. The constrain of LIDTs on the laser contrast

In laser-plasma experiments, the targets can be damaged by optical energy before the main pulses, which is generally called prepulse energy. For ultraintense femtosecond lasers, there are mainly three sources of prepulse energy. The first is ASE due to amplifier gain and incomplete Pockels cell switching. It starts a few nanoseconds before the main pulse and exists as a continuous background. The second is the ultrashort pulses with durations similar to the main pulse. They originate from the multi-reflection of the mirrors or the regenerator. The third comes from the incomplete compression due to high-order effects and spectral clipping. It forms a pedestal with a duration of a few tens of picoseconds. For simplicity, here we call them ns-ASE, fs-prepulse, and ps-pedestal, respectively. Fig.4 shows the temporal profiles of 3 ultraintense pulses from the Shanghai Superintense Ultrafast Laser Facility (SULF) at Shanghai Zhangjiang Comprehensive National Science Center [40], the CoReLS at Gwangju Insititute of Science and Technology (GIST) [41], and the CLAPA facility at Peking University. They were all measured by third-order-autocorrelators.

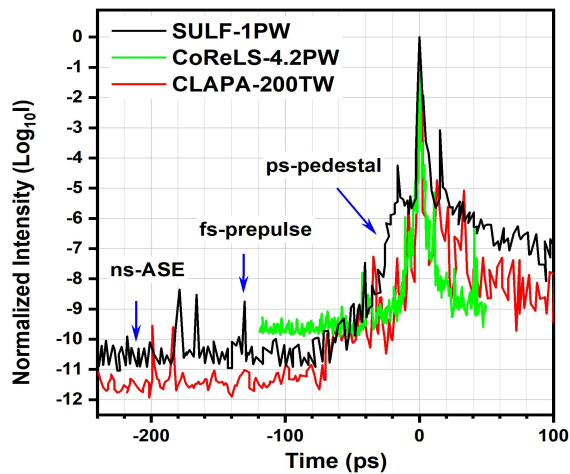

Fig.4 The temporal profiles of three ultraintense femtosecond laser pulses from the SULF, CoReLS, and CLAPA facilities, respectively.

Table 1 lists the contrast data obtained from Fig.4. The CoReLS laser has the best contrast due to the use of cross polarized wave (XPW) technique and the optical parametric amplification (OPA) front end stages. By using the double-XPW technique, the CLAPA laser has as good ps-pedestal as the CoReLS, but higher ns-ASE (which may be due to limited measurement range). The SULF has the highest ps-pedestal.

Table 1. The contrast of SULF, CoReLS, and CLAPA lasers

|  | ns-ASE@200ps | ps-pedestal@5ps | fs-prepulses |
| --- | --- | --- | --- |
| CLAPA | 1e-10 | 1e-7 | 1e-8 |
| CoReLS | 5e-11 | 1e-7 | 1e-8 |
| SULF | 5e-10 | 1e-5 | 1e-8 |

To investigate whether ultrathin targets can be safely used in the experiments performed in the above laser facilities, we assume the peak intensity in CoReLS, SULF, and CLAPA, is $5\times10^{21}$ W/cm$^2$, $1\times10^{21}$ W/cm$^2$ and $5\times10^{19}$ W/cm$^2$, respectively, based on reported results[21, 42-43]. Fig. 5 shows the DTIs of the six kinds of targets as histograms. One can easily see all the targets will be damaged by the fs-prepulses or ps-pedestal. In order to use ultrathin targets, additional contrast improvement techniques need to be applied. The most successful technique up to now, if not the only, is the plasma mirrors (PM). A plasma mirror acts on the compressed pulse as a picosecond gated temporal switch and thus reflects only the ultra-short high-intensity pulse[44], usually providing a contrast ratio improvement of the order of $10^2$ ($10^4$) for a single (double) PM. When a single PM is applied, 20%-30% laser energy will be lost as the price.

Fig. 5 can be used to evaluate how many PMs need to be applied to avoid prepulse-indeced damage and pre-expansion for a specific kind of ultrathin target. For example, by applying 1 PM to CLAPA, 2 PMs to CoReLS, and 2 PMs to SULF, the prepulse intensity will drops by more than 2/4/4 orders, respectively. The corresponding prepulse intensity histograms are drawn below the DTI histograms in Fig.5. For a specific target, if its DTIs at 50 fs/500 fs/200 ps is higher than the fs-prepulse intensity, the ps-pedestal intensity @5 ps, and the ns-ASE intensity @200 ps, respectively, such a target would still be intact until 5 ps before the main pulse. One can easily make the judgment by comparing the bars with the same color in the two histograms. The conclusion is that all the ultrathin foils can be safely used in CLAPA with a single PM and the CoReLS with double PMs, but only formvar can be safely used in SULF even with double PMs. For all the three lasers, the dominant constrain comes from the ps-pedestal intensity. When the ps-pedestal reaches the intensity of over $10^{12}$ W/cm$^2$, shock with speed about 10 nm/ps will be launched firstly in the targets[45]. Then the targets will be ionized by the fast-rising pedestal and start to expand. How such an expansion influence the ion acceleration depends on the thickness of the foil, the expansion speed, and the expansion time.

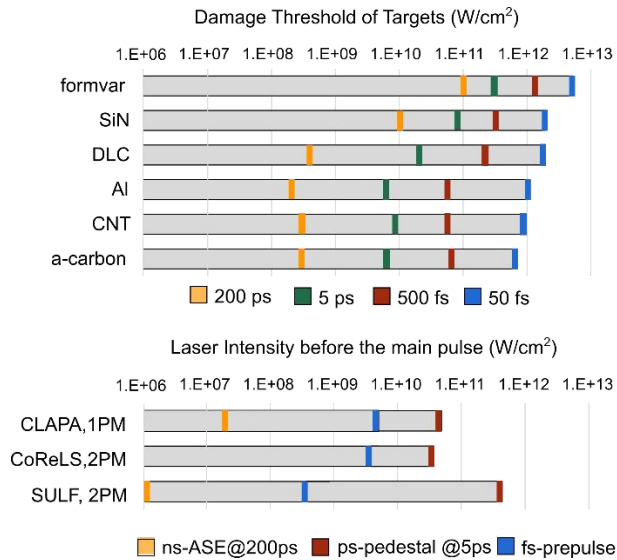

Fig.5 Comparison between DTIs of the tested ultrathin foils and the prepulse intensity.

## 6. Conclusion

In conclusion, we have performed experimental studies on the single-shot optical damage threshold of multi-type ultrathin foils used for laser-driven ion acceleration by using laser pulses with durations from 50 fs to 200 ps. We found the damage thresholds of the free-standing ultrathin foils are significantly lower than those of corresponding bulk materials. Dielectric targets such as formvar and SiN have the highest DTF among all the tested targets. The DTFs of DLC, Al, amorphous carbon, and CNT targets are all below 0.1 J/cm$^2$ with weak dependence on the pulse duration. The obtained damage thresholds are very valuable to help to judge whether a specific target can be safely used in experiments by comparing them with prepulse intensity of lasers in a simple diagram.

## Acknowledgements

This work was supported by NSFC innovation groupproject (11921006), National Grand Instrument Project(2019YFF01014402), Natural Science Foundation ofChina (Grant No. 11775010,11535001, 61631001), and State Key Laboratory Foundation of Laser Interaction with Matter (SKLLIM1806). We thank Dr. Jianhui Bin and Mr. Alinger Klaus for their helps in the experiments.